\documentstyle[sprocl,amssymb,epsf]{article}

\def\Journal#1#2#3#4{{#1} {\bf #2}, #3 (#4)}

\def\NPB{{\em Nucl. Phys.} B}
\def\PLB{{\em Phys. Lett.}  B}

\def\PRD{{\em Phys. Rev.} D}
\def\ZPC{{\em Z. Phys.} C}

\def\be{\begin{equation}}
\def\ee{\end{equation}}
\def\bea{\begin{eqnarray}}
\def\eea{\end{eqnarray}}
\makeatletter
\def\slash#1{{\mathpalette\c@ncel{#1}}} % TeXbook, bottom of p360
\makeatother
\newcommand{\deriv}{\stackrel{\leftrightarrow}{D}}
\newcommand{\derleft}{\stackrel{\leftarrow}{D}}
\newcommand{\derright}{\stackrel{\rightarrow}{D}}

\newcommand\beq{\begin{eqnarray}}
\newcommand\eeq{\end{eqnarray}}

\def\Gtilde{\tilde{G}}
\def\xslash{\rlap/{\mkern-1mu x}}

\begin{document}
\begin{flushright}
NORDITA-98/51 HE\\
hep-ph/9808229
\end{flushright}

\title{HANDBOOK OF HIGHER TWIST DISTRIBUTION AMPLITUDES OF VECTOR 
       MESONS IN QCD}

\author{PATRICIA BALL }
\address{CERN--TH, CH--1211 Gen\`eve 23, Switzerland}

\author{V.M. BRAUN }
\address{NORDITA, Blegdamsvej 17, DK--2100 Copenhagen \O,
Denmark }

\bigskip

\address{Talk presented by V.M.\ Braun at 
3rd workshop ``Continuous Advances in QCD'',
         Minneapolis, MN, USA, April 16--19,  1998.}


\maketitle\abstracts{ 
We give a summary of existing results on higher twist distribution 
amplitudes of vector mesons in QCD. Special attention is payed 
to meson mass corrections which turn out to be large. A ``shopping list'' 
is presented of most important nonperturbative parameters 
which enter distribution amplitudes.  
}

\section{General framework}

%This report attempts to give a consise summary of existing results
%on light-cone distribution amplitudes of vector mesons and is   
%motivated by the newly emerging applications of the theory of hard
%exclusive 
%processes to electroproduction and B-meson decays. 

The notion of distribution amplitudes refers to 
momentum fraction distributions of partons in the meson in a particular 
Fock state with fixed number of components. For the minimal number of 
constituents, the distribution amplitude $\phi$ is related to the 
Bethe-Salpeter wave function $\phi_{BS}$ by
\begin{equation}
  \phi(x) \sim \int^{|k_\perp| < \mu} \!\!d^2 k_\perp\,
\phi_{BS}(x,k_\perp).
\end{equation}   
The standard approach to distribution amplitudes, which is due to Brodsky
and Lepage~\cite{BLreport}, considers the hadron's parton decomposition
in the infinite momentum frame. A conceptually different, but
mathematically 
equivalent formalism is the light-cone quantization~\cite{LCQ}.
Either way, power suppressed contributions to exclusive processes in QCD,
which are commonly referred to as higher twist corrections, are thought to 
originate from three different sources:
\begin{itemize}
\item contributions of ``bad'' components in the wave function and 
      in particular of those with ``wrong'' spin projection;
\item contributions of transverse motion of quarks (antiquarks) in the 
      leading twist components, given for instance by integrals as above with 
      additional factors of $k_\perp^2$;
\item contributions of higher Fock states,  with additional gluons and/or
      quark-antiquark pairs. 
\end{itemize} 
We take a somewhat different point of view and define 
light-cone distribution amplitudes as meson-to-vacuum transition matrix 
elements of nonlocal gauge-invariant light-cone operators. This formalism 
is convenient for the study of higher-twist distributions thanks to its 
gauge and Lorentz invariance and allows to solve all equations of motion 
explicitly, relating different higher-twist distributions to each other.   
We will find that all dynamical degrees of freedom are those describing 
contributions of higher Fock states, while all other higher-twist effects 
are given in terms of the latter without any free parameters.

The report is divided into three sections, the first of which
is introductory and the last two present the summary of 
distribution amplitudes up to twist~4.    
The expressions collected in these sections  are principally the result
of recent studies reported in Refs.~\cite{BBrho,BBKT,BBS} which considerably 
extend the earlier analysis in  Ref.~\cite{CZreport}.
We use a simplified version of the set of twist-4 distributions
derived in \cite{BBS},
taking into account only contributions of the lowest conformal
partial-waves,
and for consistency discard contributions of higher partial-waves in 
twist-3 distributions in cases where they enter physical amplitudes 
multiplied by additional powers of $m_\rho$.
Four-particle distributions of twist~4 start with higher conformal spin
and must be put to zero to the present accuracy.  
The SU(3)-breaking effects are taken 
into account in leading-twist distributions and partially 
for twist-3, but neglected for twist-4. Explicit expressions are given 
for a (charged) $\rho$-meson. 
Distribution amplitudes for other vector-mesons are obtained 
by trivial substitutions.

Throughout this report we denote the meson momentum by $P_\mu$  
and introduce the light-like vectors $p$ and $z$ such that 
\begin{equation}
p_\mu = P_\mu-\frac{1}{2}z_\mu \frac{m^2_\rho}{pz}.
\label{smallp}
\end{equation}  
The meson polarization vector $e^{(\lambda)}_\mu$ is decomposed in 
projections onto the two light-like vectors and the orthogonal plane 
as
\begin{equation}
 e^{(\lambda)}_\mu = \frac{(e^{(\lambda)}\cdot z)}{pz}
\left( p_\mu -\frac{m^2_\rho}{2pz} z_\mu \right)+e^{(\lambda)}_{\perp\mu}. 
\label{polv}
\end{equation} 
We use the standard Bjorken-Drell 
convention \cite{BD65} for the metric and the Dirac matrices; in particular
$\gamma_{5} = i \gamma^{0} \gamma^{1} \gamma^{2} \gamma^{3}$,
and the Levi-Civita tensor $\epsilon_{\mu \nu \lambda \sigma}$
is defined as the totally antisymmetric tensor with $\epsilon_{0123} = 1$.
The covariant derivative is defined as 
$D_{\mu} \equiv \overrightarrow{D}_\mu= \partial_{\mu} - igA_{\mu}$, 
%which is consistent with the gauge phase factor (\ref{Pexp}), 
and we also use the notation
$\overleftarrow{D}_{\mu} = \overleftarrow\partial_{\mu} 
+ig A_{\mu}$ in later sections. The dual gluon field strength
tensor is defined as $\widetilde{G}_{\mu\nu} =
\frac{1}{2}\epsilon_{\mu\nu \rho\sigma} G^{\rho\sigma}$.

\subsection{Conformal partial wave expansion}

Conformal partial wave expansion in 
QCD~\cite{BL80,Makeenko,O82,Frishman,BF90,Mueller}
parallels the partial wave expansion
of wave functions in standard quantum mechanics,
which allows to separate the dependence on angular 
coordinates from radial ones. The basic idea is to write down 
distribution amplitudes as a sum of contributions from different 
conformal spins. For a given spin, the dependence on the momentum 
fractions is fixed by the symmetry. To specify the function, 
one has to fix the coefficients in this expansion at some scale;
 conformal invariance of the QCD Lagrangian then guarantees that there 
is no mixing between contributions of different spin to leading logarithmic 
accuracy. For leading twist distributions the mixing matrix becomes 
diagonal in the conformal basis and
the anomalous dimensions are ordered with spin.  
Thus, only the first few ``harmonics'' contribute at sufficiently large scales 
(for sufficiently hard processes).
  
For higher twist distributions, the use of the conformal basis offers 
a crucial  advantage of ``diagonalizing'' the equations of motion:
since conformal transformations commute
with the QCD equations of motion, the corresponding constraints
 can be solved order by order in the conformal expansion.
Note that relations between different distributions obtained in this way
are exact: despite the fact that conformal symmetry is broken by quantum 
corrections, equations of motion are not renormalized and remain the same as 
in free theory.  

The general procedure to construct the conformal expansion
for arbitrary multiparticle distributions was developed in \cite{O82,BF90}.
To this end each constituent field has to be decomposed (using projection 
operators, if necessary) in components  with fixed (Lorentz) spin 
projection onto the light-cone.
 
Each such component corresponds to a so-called quasiprimary field in 
the language of 
conformal field theories, and has conformal spin 
\begin{equation}
  j=\frac{1}{2}\,(l+s),
\label{eq:cspin}
\end{equation}
where $l$ is the canonical dimension  and $s$ the (Lorentz) spin 
projection.\footnote{$l=3/2$ for quarks and $l=2$ for gluons; 
the quark field is decomposed as $\psi = \psi_++\psi_- \equiv 
(1/2)\!\not\!\!z\!\!\not\!\!p\psi+
(1/2)\!\not\!\!p\!\!\not\!\!z\psi$ with 
spin projections $s=+1/2$ and $s=-1/2$, respectively. For the gluon 
field strength there are three possibilities:
 $z_\mu G_{\mu\perp}$ corresponds to $s=+1$, 
$p_\mu G_{\mu\perp}$ has $s=-1$ and
$G_{\perp\perp}$, $z_\mu p_\nu G_{\mu\nu}$ correspond both to $s=0$.}

Multi-particle states built of quasiprimary fields can be expanded in 
irreducible unitary representations with increasing conformal spin.
An explicit expression for the distribution amplitude 
of a multi-particle state with the lowest conformal spin 
 $j=j_1+\ldots+j_m$ built of $m$ primary fields with the spins $j_k$ is
\begin{equation}
\phi_{as}(\alpha_1,\alpha_2,\ldots,\alpha_m) = 
\frac{\Gamma[2j_1+\ldots +2j_m]}{\Gamma[2j_1]\ldots \Gamma[2j_m]}
\alpha_1^{2j_1-1}\alpha_2^{2j_2-1}\ldots \alpha_m^{2j_m-1}.
\label{eq:asymptotic}
\end{equation}
Here $\alpha_k$ are the corresponding momentum fractions.
This state is nondegenerate and cannot mix with 
other states because of conformal symmetry.
Multi-particle irreducible representations with higher spin
$j+n,n=1,2,\ldots$, 
are given by  polynomials of $m$ variables (with the constraint 
$\sum_{k=1}^m \alpha_k=1$ ) which are orthogonal over
 the weight function (\ref{eq:asymptotic}).

 \subsection{Equations of motion}  

We collect here exact operator identities which can be derived using the 
approach of \cite{BB89} and which present a nonlocal equivalent
of the equations of motion for Wilson local operators. 
Taking suitable matrix elements
one derives a set of relations between distribution amplitudes, which 
generally allow to express two-particle distributions of higher twist 
in terms of three-particle distributions. 
The corresponding relations 
are taken into account in the results given in later sections. 
The rationale for presenting the operator relations themselves 
is that we found them useful in practical calculations: 
\begin{eqnarray}
\partial^\mu \bar u(x) \gamma_\mu d(-x) & = & -i\int_{-1}^1 dv\, \bar
u(x) x^\nu gG_{\nu \mu}(vx) \gamma^\mu d(-x),\\
\frac{\partial\phantom{x_\mu}}{\partial x_\mu} \,\bar u(x)
\gamma_{\mu} d(-x) & = & -i\int_{-1}^1 dv\, v \bar
u(x) x^\nu gG_{\nu \mu}(vx) \gamma^\mu d(-x),\label{eq:sec}
\end{eqnarray}
\begin{eqnarray}
\partial^\mu \bar u(x) \sigma_{\mu\nu} d(-x) & = &
-i\,\frac{\partial\phantom{x_\nu}}{\partial x_\nu} \,\bar u(x)
d(-x) + \int_{-1}^1 dv\, v \bar u(x)
x^\rho gG_{\rho\nu}(vx)d(-x)\nonumber\\
& & {} -i\int_{-1}^1 dv\, \bar u(x) x^\rho gG_{\rho\mu}(vx) 
\sigma^\mu_{\phantom{\mu}\nu} d(-x),\label{eq:rel1}\\
\frac{\partial\phantom{x_\nu}}{\partial x_\mu} \,\bar u(x)
\sigma_{\mu\nu} d(-x) & = & -i\partial_\nu \bar u(x) d(-x) + 
\int_{-1}^1 dv\, \bar u(x) x^\rho gG_{\rho\nu}(vx)d(-x)\nonumber\\
& & {} -
i\int_{-1}^1 dv\, v \bar u(x) x^\rho gG_{\rho\mu}(vx) 
\sigma^\mu_{\phantom{\mu}\nu} d(-x),
\label{eq:rel2}
\end{eqnarray}
\begin{eqnarray}
%\frac{\partial}{\partial x_{\mu}}
%\bar{u}(x) \sigma_{\mu \nu} x^{\nu} 
%d (-x)  &=&
%i \int_{-1}^{1}\! dv\, v \; \bar{u}(x) x^{\alpha}
%\sigma_{\alpha \beta} x^{\mu}gG_{\mu \beta}(vx) d(-x)
%\nonumber \\
%&-& i x^{\beta}\partial_{\beta} \left\{ \bar{u}(x)
% d(-x) \right\} - (m_{u} - m_{d}) \bar{u}(x)
%\!\not\!x \,d(-x),
%\label{eq:3id1} \\
\bar{u}(x)d(-x) - \bar{u}(0) d(0) &=&
\int_{0}^{1} dt \int_{-t}^{t}dv\,
\bar{u}(tx) x^{\alpha} \sigma_{\alpha \beta}
x^{\mu} gG_{\mu \beta} (vx)  d(-tx)
\nonumber\\
&&{}+ i \int_{0}^{1}\!dt\, \partial^{\alpha}\left\{
\bar{u}(tx) \sigma_{\alpha \beta}x^{\beta}d(-tx) \right\}.
%\nonumber\\&&
%+ i (m_{u} + m_{d}) \int_{0}^{1} \!dt \,\bar{u}(tx)
%\!\not\! x \, d(-tx).
\label{eq:3id2}
\end{eqnarray}
In all cases gauge factors are implied in between the constituent fields,
\begin{equation}
[x,y] =\mbox{\rm Pexp}[ig\!\!\int_0^1\!\! dt\,(x-y)_\mu A^\mu(tx+(1-t)y)],
\label{Pexp}
\end{equation}
and we introduced  a shorthand notation
for the derivative over the total translation:
\begin{equation}
\partial_{\alpha}\left\{ \bar{u}(tx)
\Gamma  d(-tx) \right\} \equiv
\left. \frac{\partial}{\partial y^{\alpha}}
\left\{ \bar{u}(tx + y) \Gamma d(-tx + y)\right\} 
\right|_{y \rightarrow 0}
\label{eq:3tdrv}
\end{equation}
with the generic Dirac matrix structure $\Gamma$.
For simplicity, we omit
operators involving quark masses, see~\cite{BBKT,BBS}.  

Two more relations are:
\beq
\lefteqn{
\bar{u}(x)\gamma_\mu d(-x) =\int_0^1\, dt\, 
{ \partial \over \partial x_\mu}\,\bar{u}(tx)\xslash  
d(-tx)}\nonumber\\
&&{}- \int_0^1\!dt\,\int_{-t}^t\!dv\,\bar{u}(tx)\Big\{t g\Gtilde_{\mu\nu}(vx)
x^\nu \xslash \gamma_5 +ivgG_{\mu\nu}(vx)x^\nu \xslash \Big\}
d(-tx)\nonumber\\
&&{}+ \int_0^1\!dt\,t\,\int_{-t}^t\!dv\,\bar{u}(tx)\Big\{
x^2 g\Gtilde_{\mu\nu}(vx) \gamma^\nu \gamma_5 
-x_\mu x^\nu g\tilde G_{\nu\rho}(vx)\gamma^\rho\gamma_5\Big\}
d(-tx)   
\nonumber\\
&&{}
-i\epsilon_{\mu\nu\alpha\beta}\!\int_0^1\!\!\!dt\,t\,x^\nu\partial^\alpha
\left[ \bar{u}(tx)\gamma^\beta\gamma_5 d(-tx)\right],
%+ (m_u\! -\!m_d) x^\nu\!\!\!
% \int_0^1\!\!\!dt\,t\,\bar{u}(tx)\sigma_{\nu\mu}d(-tx),
\label{eq4.1}
\eeq
and similarly with an additional $\gamma_5$:
\beq
\lefteqn{
\bar{u}(x)\gamma_\mu\gamma_5 d(-x) =\int_0^1\, dt\, 
{ \partial \over \partial x_\mu}\,\bar{u}(tx)\xslash\gamma_5  
d(-tx)}\nonumber\\
&&{}- \int_0^1\!dt\,\int_{-t}^t\!dv\,\bar{u}(tx)\Big\{t g\Gtilde_{\mu\nu}(vx)
x^\nu \xslash  +ivgG_{\mu\nu}(vx)x^\nu \xslash\gamma_5 \Big\}
d(-tx)\nonumber\\
&&{}+ \int_0^1\!dt\,t\,\int_{-t}^t\!dv\,\bar{u}(tx)\Big\{
x^2 g\Gtilde_{\mu\nu}(vx) \gamma^\nu  
-x_\mu x^\nu g\tilde G_{\nu\rho}(vx)\gamma^\rho\Big\}
d(-tx)   
\nonumber\\
&&{}
-i\epsilon_{\mu\nu\alpha\beta}\!\int_0^1\!\!\!dt\,t\,x^\nu\partial^\alpha
\left[ \bar{u}(tx)\gamma^\beta d(-tx)\right].
%+ (m_u\! +\!m_d) x^\nu\!\!\!
% \int_0^1\!\!\!dt\,t\,\bar{u}(tx)\sigma_{\nu\mu}\gamma_5d(-tx).
\label{eq4.2}
\eeq
Finally, the following formula is sometimes useful:
\begin{eqnarray}
\lefteqn{
\frac{\partial^2}{\partial x_\alpha \partial x^\alpha} \bar u(x)\Gamma\, d(-x)
 = -\partial^2 \bar u(x)\Gamma\, d(-x) +\bar u(x) 
[\Gamma\sigma G+\sigma G\,\Gamma]d(-x)}
\nonumber\\
&&{}-2ix^\nu \frac{\partial}{\partial x_\mu}\!\int_{-1}^1\!\!dv\,v\,\bar u(x)
\Gamma G_{\nu\mu}(vx)d(-x)
-2ix^\nu \partial_\mu\! \int_{-1}^1\!\!dv\,\bar u(x)
\Gamma G_{\nu\mu}(vx)d(-x)
\nonumber\\
&&{}+2\int_{-1}^1\!dv\int_{-1}^v\!dt\,(1+vt)\bar u(x)\Gamma x^\mu x^\nu
G_{\mu\rho}(vx)G^{\rho}_{\phantom{\nu}\nu}(tx)d(-x)
\nonumber\\
&&{}+ix^\nu \int_{-1}^1\!\!dv\,(1+v^2)\,\bar u(x)\Gamma 
[D_\mu,G^{\mu}_{\phantom{\nu}\nu}](vx) d(-x)
\label{twoderiv}
\end{eqnarray}
where 
$[D_\mu,G^{\mu}_{\phantom{\nu}\nu}] =-t^A(\bar\psi\gamma_\nu t^A\psi)$ 
assuming  summation over light flavors $\psi$.

\subsection{Meson mass corrections} 
The structure of meson mass corrections in exclusive processes is 
in general more complicated than of target mass corrections in 
deep inelastic scattering in which case they can be resummed using the
Nachtmann variable \cite{N73}. For illustration, consider the simplest 
matrix element
\begin{eqnarray}
\lefteqn{
 \langle0|\bar u(x)\xslash d(-x)|\rho^-(P,\lambda)\rangle =}\makebox[1cm]{\ }
\nonumber\\&=& f_\rho m_\rho
  (e^{(\lambda)}x)\int_0^1 \!du\, e^{i(2u-1)Px}
  \Big[\phi(u)+\frac{x^2}{4}\Phi(u)+O(x^4)\Big].
\label{mass1}
\end{eqnarray}
We assume that $x^2 \ll \Lambda_{\rm QCD}^{-2}$, but nonzero, $\phi(u)$ 
is the twist-2 distribution amplitude and $\Phi(u)$ describes
higher-twist corrections in which we want to calculate  
``kinematic'' contributions due to nonzero $\rho$-meson mass. 

A common wisdom tells that meson mass corrections are related to contributions
of leading twist operators. Indeed, conditions of symmetry and zero traces
for twist-2 local operators imply 
\begin{eqnarray}
\lefteqn{
\langle0|\Big[\bar u\xslash (i\deriv x)^n d\Big]_{\rm tw.2}
|\rho^-(P,\lambda)\rangle = }\makebox[1cm]{\ }
\nonumber\\&=&f_\rho m_\rho
(e^{(\lambda)}x)\Bigg[(Px)^n-\frac{x^2m_\rho^2}{4}\frac{n(n-1)}{n+1}(Px)^{n-2}
\Bigg]\langle\!\langle O_n\rangle\!\rangle,
\label{mass2}
\end{eqnarray} 
where $[\ldots]_{\rm tw.2}$ denotes taking the leading twist part (subtraction
of traces, in this case) and $\langle\!\langle O_n\rangle\!\rangle$ is
the reduced matrix element related to the $n$-th moment of the leading twist 
distribution 
\begin{equation}
  M_n^{(\phi)}\equiv \int_0^1 \!du\,(2u-1)^n\phi(u) = 
  \langle\!\langle O_n\rangle\!\rangle.
\end{equation}
Expanding (\ref{mass1}) at short distances $x\rightarrow0$ and comparing with 
(\ref{mass2}), we find that the same reduced matrix element gives a 
contribution to the twist~4 distribution amplitude
\begin{equation}
  M_n^{(\Phi)}\equiv \int_0^1 \!du\,(2u-1)^n\Phi(u) = 
  \frac{1}{n+3}m_\rho^2\langle\!\langle O_{n+2}\rangle\!\rangle,
\label{mass3}
\end{equation}
which is the direct analogue of Nachtmann's correction for deep inelastic
scattering. 

The result in (\ref{mass3}) is, however, incomplete. The reason is that 
in exclusive processes one has to take into account higher-twist operators 
containing full derivatives, and vacuum-to-meson matrix elements of 
such operators reduce, in certain cases, to powers of the meson mass times
reduced matrix elements of leading twist operators. 
For the case at hand, write\cite{BB89}
\begin{eqnarray}
  \bar u(x)\xslash d(-x) &=& \Big[\bar u(x)\xslash d(-x)\Big]_{\rm tw.2}+
\frac{x^2}{4}\int_0^1\!\!dt\,
\frac{\partial^2}{\partial x_\alpha \partial x^\alpha}\bar u(tx)\xslash d(-tx)
+O(x^4)
\nonumber\\
&=& \Big[\bar u(x)\xslash d(-x)\Big]_{\rm tw.2} -
\frac{x^2}{4}\int_0^1\!\!dt\,t^2\,\partial^2[\bar u(tx)\xslash d(-tx)]
\nonumber\\
&&+{\rm ~contributions~of~operators~with~gluons}+O(x^4)
% \Big[\bar u(x)\xslash d(-x)\Big]_{\rm tw.4} + \ldots
%O(\mbox{\rm tw.6})
\end{eqnarray} 
where we used Eq.~(\ref{twoderiv}) to arrive at the last line.
In the matrix element one can make the substitution 
$\partial^2\rightarrow -m_\rho^2$.
Expanding, again, at short distances, and comparing with the similar 
expansion of (\ref{mass1}) we get an additional 
contribution to $M_n^{(\Phi)}$ so that the corrected version of 
(\ref{mass3}) becomes 
\begin{equation}
  M_n^{(\Phi)} = 
  \frac{1}{n+3}m_\rho^2\big[\langle\!\langle O_{n+2}\rangle\!\rangle
    +\langle\!\langle O_{n}\rangle\!\rangle\big]+{\rm ~gluons}.
\label{mass4}
\end{equation}
Assuming the asymtotic form of the leading-twist 
distribution amplitude $\phi$,
$\phi(u)=6u(1-u)$, so that  
$\langle\!\langle O_{n}\rangle\!\rangle =3/[2(n+1)(n+3)]$, this equation 
for moments is easily solved and gives
\begin{equation}
  \Phi(u) = 30 u^2(1-u)^2
  \left[\frac{2}{5}m_\rho^2 +\frac{4}{3}m^2_\rho\zeta_4\right],
\label{mass5}
\end{equation}
where we have included the ``genuine'' twist~4 correction (term in $\zeta_4$)
due to the twist~4 quark-gluon operator, see definition in 
Eq.~(\ref{def:zeta34}). 
The QCD sum rule estimate is $\zeta_4\sim 0.15$~\cite{BK86}, so that the meson 
mass effect onto the twist~4 distribution function 
 is by a factor two larger than the ``genuine'' twist~4 correction. 
This is an important difference to deep inelastic scattering, where the 
target mass corrections are small.

The present discussion is still oversimplified and does not provide with 
a complete separation of meson mass effects. The major complication 
arises because of  contributions of operators of the type
\begin{equation}
\partial_{\mu_1} \Big[\bar u\gamma_{\mu_1}(i\!\deriv_{\mu_2})
\ldots(i\!\deriv_{\mu_n})d\Big]_{\rm tw.2}.
\end{equation}
Such operators can be expressed in terms of operators with extra gluon fields,
which means, conversely, that certain combinations of $\bar qGq$ operators 
reduce to divergences of leading twist operators and give rize to extra 
meson mass correction terms. The corresponding corrections to twist~4 
distributions involve, however, 
 higher-order contributions in the conformal expansion of the distribution 
amplitudes of leading twist and do not affect the result in (\ref{mass5}),
which is to  leading conformal spin accuracy.
A detailed analysis will be presented in~\cite{BBS}.

\section{Summary of chiral-even distributions}
Two-particle quark-antiquark distribution amplitudes are defined 
as matrix elements of non-local operators on the light-cone \cite{BBKT}:
\begin{eqnarray}
\lefteqn{\langle 0|\bar u(z) \gamma_{\mu} d(-z)|\rho^-(P,\lambda)\rangle 
 = f_{\rho} m_{\rho} \left[ p_{\mu}
\frac{e^{(\lambda)}\cdot z}{p \cdot z}
\int_{0}^{1} \!du\, e^{i \xi p \cdot z} \phi_{\parallel}(u, \mu^{2}) \right. 
}\hspace*{0.6cm}\nonumber \\
&&{}\left.+ e^{(\lambda)}_{\perp \mu}
\int_{0}^{1} \!du\, e^{i \xi p \cdot z} g_{\perp}^{(v)}(u, \mu^{2}) 
%\nonumber \\
- \frac{1}{2}z_{\mu}
\frac{e^{(\lambda)}\cdot z }{(p \cdot z)^{2}} m_{\rho}^{2}
\int_{0}^{1} \!du\, e^{i \xi p \cdot z} g_{3}(u, \mu^{2})
\right]\hspace*{0.4cm}
\label{eq:vda}
\end{eqnarray}
and 
\begin{eqnarray}
\lefteqn{\langle 0|\bar u(z) \gamma_{\mu} \gamma_{5} 
d(-z)|\rho^-(P,\lambda)\rangle=}\hspace*{0.6cm}\nonumber\\ 
&=& \frac{1}{2}\left(f_{\rho} - f_{\rho}^{T}
\frac{m_{u} + m_{d}}{m_{\rho}}\right)
m_{\rho} \epsilon_{\mu}^{\phantom{\mu}\nu \alpha \beta}
e^{(\lambda)}_{\perp \nu} p_{\alpha} z_{\beta}
\int_{0}^{1} \!du\, e^{i \xi p \cdot z} g^{(a)}_{\perp}(u, \mu^{2}).
\label{eq:avda}
\end{eqnarray}
 For brevity, here and below we do not show the gauge factors between the 
quark and the antiquark fields and use the short--hand notation
$$\xi = u - (1-u) = 2u-1.$$ 
The vector and tensor  decay constants $f_\rho$ and $f_\rho^T$ are defined,
as usual, as
\begin{eqnarray}
\langle 0|\bar u(0) \gamma_{\mu}
d(0)|\rho^-(P,\lambda)\rangle & = & f_{\rho}m_{\rho}
e^{(\lambda)}_{\mu},
\label{eq:fr}\\
\langle 0|\bar u(0) \sigma_{\mu \nu} 
d(0)|\rho^-(P,\lambda)\rangle &=& i f_{\rho}^{T}
(e_{\mu}^{(\lambda)}P_{\nu} - e_{\nu}^{(\lambda)}P_{\mu}).
\label{eq:frp}
\end{eqnarray}
The distribution amplitude $\phi_\parallel$ is of twist-2,
$g_\perp^{(v)}$ and $g_\perp^{(a)}$ are twist-3 and $g_3$ is twist-4. 
All four functions $\phi=\{\phi_\parallel,
g_\perp^{(v)},g_\perp^{(a)},g_3\}$ are normalized as
\begin{equation}
\int_0^1\!du\, \phi(u) =1,
\label{eq:norm}
\end{equation}
which can be checked by comparing the two sides
of the  defining equations in the limit $z_\mu\to 0$ and using the
equations of motion.
We keep the (tiny) corrections proportional 
to the $u$ and $d$ quark-masses $m_u$ and $m_d$ to indicate the 
SU(3)-breaking corrections for $K^*$- and $\phi$-mesons.

In addition, we have to define three-particle distributions:
\begin{eqnarray}
\lefteqn{\langle 0|\bar u(z) g\widetilde G_{\mu\nu}(vz)\gamma_\alpha\gamma_5 
  d(-z)|\rho^-(P,\lambda)\rangle =
  f_\rho m_\rho p_\alpha[p_\nu e^{(\lambda)}_{\perp\mu}
 -p_\mu e^{(\lambda)}_{\perp\nu}]{\cal A}(v,pz)}\hspace*{4cm}
\nonumber\\ &
+ & f_\rho m_\rho^3\frac{e^{(\lambda)}\cdot z}{pz}
[p_\mu g^\perp_{\alpha\nu}-p_\nu g^\perp_{\alpha\mu}] \widetilde\Phi(v,pz)
\nonumber\\& + &
 f_\rho m_\rho^3\frac{e^{(\lambda)}\cdot z}{(pz)^2}
p_\alpha [p_\mu z_\nu - p_\nu z_\mu] \widetilde\Psi(v,pz),\hspace*{0.4cm}\\
%\end{eqnarray}
%\begin{eqnarray}
\lefteqn{\langle 0|\bar u(z) g G_{\mu\nu}(vz)i\gamma_\alpha 
  d(-z)|\rho^-(P)\rangle =
  f_\rho m_\rho p_\alpha[p_\nu e^{(\lambda)}_{\perp\mu} 
  - p_\mu e^{(\lambda)}_{\perp\nu}]{\cal V}(v,pz)}\hspace*{4cm}
\nonumber\\&
+& f_\rho m_\rho^3\frac{e^{(\lambda)}\cdot z}{pz}
[p_\mu g^\perp_{\alpha\nu} - p_\nu g^\perp_{\alpha\mu}] \Phi(v,pz)
\nonumber\\&
+ & f_\rho m_\rho^3\frac{e^{(\lambda)}\cdot z}{(pz)^2}
p_\alpha [p_\mu z_\nu - p_\nu z_\mu] \Psi(v,pz),
\end{eqnarray}
where 
\begin{equation}
   {\cal A}(v,pz) =\int {\cal D}\underline{\alpha} 
e^{-ipz(\alpha_u-\alpha_d+v\alpha_g)}{\cal A}(\underline{\alpha}),
\end{equation}
etc., and $\underline{\alpha}$ is the set of three momentum fractions
$\underline{\alpha}=\{\alpha_d,\alpha_u,\alpha_g\}$.
 The integration measure is defined as 
\begin{equation}
 \int {\cal D}\underline{\alpha} \equiv \int_0^1 d\alpha_d
  \int_0^1 d\alpha_u\int_0^1 d\alpha_g \,\delta(1-\sum \alpha_i).
\label{eq:measure}
\end{equation}
The distribution amplitudes ${\cal V}$ and ${\cal A}$ are of twist-3,
while the rest is twist-4 and we have not shown further Lorentz structures 
corresponding to twist-5 contributions\footnote{We use  
a different normalization of three-particle twist-3 distributions
compared to \cite{BBKT}.}.

Calculation of exclusive amplitudes involving a large momentum-transfer
reduces to evaluation of meson-to-vacuum transition matrix elements
of non-local operators, which can be expanded in powers of the deviation
from the light-cone.
To twist-4 accuracy one can use the expression for the axial-vector matrix 
element in (\ref{eq:avda}) as it stands, replacing the light-cone 
vector $z_\mu$ by the actual quark-antiquark separation $x_\mu$.
For the vector operator, the light-cone expansion to twist-4 accuracy 
reads:
\begin{eqnarray}
\lefteqn{\langle 0|\bar u(x) \gamma_\mu d(-x)|\rho^-(P,\lambda)\rangle 
 =}\hspace*{2cm}\nonumber\\
&= & f_\rho m_\rho \Bigg\{
\frac{e^{(\lambda)}x}{Px}\, P_\mu \int_0^1 du \,e^{i\xi Px}
\Big[\phi_\parallel(u,\mu)
+\frac{m^2_\rho x^2}{4}  {\Bbb A}(u,\mu)\Big]
\nonumber\\
&&{}+\left(e^{(\lambda)}_\mu-P_\mu\frac{e^{(\lambda)}x}{Px}\right)
\int_0^1 du\, e^{i\xi Px} \,g^{(v)}_\perp(u,\mu)
\nonumber\\
&&{}-\frac{1}{2}x_\mu \frac{e^{(\lambda)}x}{(Px)^2} m^2_\rho \int_0^1 du 
\, e^{i\xi Px} {\Bbb C} (u,\mu)
\Bigg\},
\label{eq:OPE1}
\end{eqnarray}
where
\begin{equation}
  {\Bbb C}(u) = g_3(u)+\phi_\parallel(u) -2 g^{(v)}_\perp(u),
\end{equation}
and ${\Bbb A}(u)$ can be related to integrals of three-particle distributions 
using the equations of motion. All distribution functions in (\ref{eq:OPE1})
are assumed to be normalized at the scale $\mu^2\sim x^{-2}$ (to 
leading-logarithmic accuracy). 
%In practical calculations it is sometimes
%convenient to use integrated distributions
%\begin{equation}
% {\Bbb C}^{(i)}(u) = -\int_0^u\!dv\,{\Bbb C}(v),\qquad 
% {\Bbb C}^{(ii)}(u) = -\int_0^u\!dv\,{\Bbb C}^{(i)}(v).
%\label{eq:c12}
%\end{equation}

For the leading twist-2 distribution amplitude $\phi_\parallel$ we use
\begin{equation}\label{eq:phipar}
\phi_\parallel(u) =  6 u\bar u \left[ 1 + 3 a_1^\parallel\, \xi +
a_2^\parallel\, \frac{3}{2} ( 5\xi^2  - 1 ) \right].
\end{equation}
The parameters $a_{1,2}^\parallel$ are defined as the local matrix
elements
\begin{eqnarray}
\langle 0|\bar u \slash{z} (i\deriv z)  d|\rho^-(P,\lambda)\rangle &
= & (e^{(\lambda)}z) (pz) f_\rho m_\rho \,\frac{3}{5}\, a_1^\parallel,
\nonumber\\
\langle 0|\bar u \slash{z} (i\deriv z)^2  d|\rho^-(P,\lambda)\rangle &
= & (e^{(\lambda)} z) (p z)^2 f_\rho m_\rho \left\{\frac{1}{5} + 
\frac{12}{35}\, a_2^\parallel \right\}.\label{eq:a12}
\end{eqnarray}
The numerical values are specified in Table~\ref{tab:para}.
The expressions for higher-twist distributions given below correspond
to the simplest self-consistent approximation that satisfies the  
QCD equations of motion \cite{BBKT,BBS}:
\begin{itemize}
\item{} Three-particle distributions of twist-3:
\begin{eqnarray}
{\cal V} (\underline{\alpha}) &=& 
540\, \zeta_3 \omega^V_3 (\alpha_d-\alpha_u)\alpha_d \alpha_u\alpha_g^2,
\label{modelV}\\
{\cal A} (\underline{\alpha}) &=& 
360\,\zeta_3 \alpha_d \alpha_u \alpha_g^2 
\Big[ 1+ \omega^A_{3}\frac{1}{2}(7\alpha_g-3)].
\label{modelA}
 \end{eqnarray}
\item{} Two-particle distributions of twist-3:
\begin{eqnarray}
g_\perp^{(a)}(u) & = & 6 u \bar u \!\left[ 1 + a_1^\parallel \xi +
\!\left\{\frac{1}{4}a_2^\parallel +
\frac{5}{3}\, \zeta_{3} \!\left(1-\frac{3}{16}\,
\omega^A_{3}+\frac{9}{16}\omega^V_3\!\right)\!\right\}
(5\xi^2-1)\right]\nonumber\\
& & {} + 6\, \widetilde{\delta}_+ \,  (3u \bar u + \bar u \ln \bar u +
u \ln u ) + 
6\, \widetilde{\delta}_- \,  (\bar u \ln \bar u - u \ln u),\\
 g_\perp^{(v)}(u) & = & \frac{3}{4}(1+\xi^2)
+ a_1^\parallel\,\frac{3}{2}\, \xi^3 
 + \left(\frac{3}{7} \, 
a_2^\parallel + 5 \zeta_{3} \right) \left(3\xi^2-1\right)
 \nonumber\\
& & {}+ \left[ \frac{9}{112}\, a_2^\parallel 
+ \frac{15}{64}\, \zeta_{3}\Big(3\,\omega_{3}^V-\omega_{3}^A\Big)
 \right] \left( 3 - 30 \xi^2 + 35\xi^4\right)\nonumber\\
& & {}+\frac{3}{2}\,\widetilde{\delta}_+\,(2+\ln u + \ln\bar u) +
\frac{3}{2}\,\widetilde{\delta}_-\, ( 2 \xi + \ln\bar u - \ln u).\label{eq:gv}
\end{eqnarray}
\item{} Three-particle distributions of twist-4:
\begin{eqnarray}
\widetilde \Phi (\underline{\alpha}) &=& 
  \Big[-\frac{1}{3}\zeta_{3}+\frac{1}{3}\zeta_{4}\Big] 
   30(1-\alpha_g)\alpha_g^2,
\nonumber\\
  \Phi (\underline{\alpha}) &=& 
  \Big[-\frac{1}{3}\zeta_{3}+\frac{1}{3}\zeta_{4}\Big] 
   30(\alpha_u-\alpha_d)\alpha_g^2,
\nonumber\\
  \widetilde\Psi (\underline{\alpha}) &=& 
  \Big[\frac{2}{3}\zeta_{3}+\frac{1}{3}\zeta_{4}\Big] 
   120 \alpha_u\alpha_d\alpha_g,
\nonumber\\
 \Psi (\underline{\alpha}) &=& 0.
\end{eqnarray}
\item{} Two-particle distributions of twist-4: 
\begin{eqnarray}
%\chi_{as}(u) &=& \Bigg[\frac{2}{5}+\frac{4}{3} d_{4\rho}\Bigg]
%                  30 u^2(1-u)^2
%\nonumber\\
{\Bbb A}(u) &=& \Bigg[\frac{4}{5}+ \frac{4}{105}\, a_2^\parallel+ 
\frac{20}{9} \zeta_{4}
                    +\frac{8}{9} \zeta_{3}\Bigg]30 u^2(1-u)^2,
\nonumber\\
g_3(u) &=& 6u(1-u) + \Bigg[\frac{1}{7}\,a_2^\parallel + \frac{10}{3} \zeta_{4}
                    -\frac{20}{3} \zeta_{3}\Bigg](1-3 \xi^2),
\nonumber\\
{\Bbb C}(u) &=& \Bigg[\frac{3}{2}-\frac{2}{7}\,a_2^\parallel+
                     \frac{10}{3} \zeta_{4}
                    +\frac{10}{3} \zeta_{3}\Bigg](1-3 \xi^2),
%\nonumber\\
% {\Bbb C}^{(ii)}(u) &=& \Bigg[\frac{3}{2}+\frac{10}{3} \zeta_{4}
%                    +\frac{10}{3} \zeta_{3}\Bigg]u^2(1-u)^2,
\end{eqnarray}
\end{itemize} 
where the dimensionless couplings $\zeta_3$ and $\zeta_4$ are 
defined as local matrix elements
\begin{eqnarray}
\lefteqn{\langle0|\bar u g\tilde G_{\mu\nu}\gamma_\alpha
 \gamma_5 d|\rho^-(P,\lambda)\rangle = 
f_\rho m_\rho \zeta_{3}
\Bigg[
e^{(\lambda)}_\mu\Big(P_\alpha P_\nu-\frac{1}{3}m^2_\rho \,g_{\alpha\nu}\Big)  
}\hspace*{2cm}\nonumber\\
& & 
-e^{(\lambda)}_\nu\Big(P_\alpha P_\mu-\frac{1}{3}m^2_\rho \,g_{\alpha\mu}\Big)
\Bigg]
%\nonumber\\
+ \frac{1}{3}f_\rho m_\rho^3 \zeta_{4}
\Bigg[e^{(\lambda)}_\mu g_{\alpha\nu}- e^{(\lambda)}_\nu g_{\alpha\mu}\Bigg],  
\hspace*{0.6cm}\label{def:zeta34}\end{eqnarray}
and have been estimated from QCD sum-rules 
 \cite{ZZC85,BK86}. $\omega_3^V$ is defined as
\begin{eqnarray}
\lefteqn{\langle 0|\bar u \slash{z} (gG_{\alpha\beta} 
z^\alpha i(\derright z) - (i\derleft z) g G_{\alpha\beta}z^\alpha) 
d |\rho^-(P,\lambda)\rangle  =}\hspace*{4cm}\nonumber\\
& = & i(pz)^3 
e^{(\lambda)}_{\perp\beta}
m_\rho f_\rho \frac{3}{28}\, \zeta_3 \omega_3^V + O(z_\beta)
\end{eqnarray}
and $\omega_3^A$ is defined as
\begin{equation}
\langle 0|\bar u \slash{z}\gamma_5 \left[ i Dz,g\tilde{G}_{\mu\nu}
  z^\mu \right] d |\rho^-(P,\lambda)\rangle = -(pz)^3 
e^{(\lambda)}_{\perp\nu} m_\rho f_\rho \zeta_3 \left( \frac{3}{7} +
  \frac{3}{28} \omega_3^A\right) + O(z_\beta).
\end{equation}
The terms in $\delta_\pm$ and $\widetilde\delta_\pm$ specify quark-mass 
corrections in twist-3 distributions induced by the equations of motion.
The numerical values of these and other coefficients are listed in 
Tables~\ref{tab:para} and \ref{tab:para2}\footnote{In the 
notation of Ref.~\cite{BBKT},
$\omega_{1,0}^A\equiv \omega_3^A$, $ \zeta_3^A\equiv \zeta_3$, and 
$\zeta_3^V \equiv (3/28)\zeta_3\omega_3^V$.}.
Note that we neglect SU(3)-breaking effects in twist-4 distributions
and in gluonic parts of twist-3 distributions. In Fig.~1 we plot the
leading-twist distribution amplitude $\phi_\parallel$ for $\rho$,
$K^*$ and $\phi$ mesons.
\begin{figure}
\caption[]{The twist-2 distribution amplitude $\phi_\parallel(u)$ for
  $\rho$, $K^*$ and $\phi$ mesons. Renormalization point is $\mu =
  1\,$GeV.}
\centerline{\epsffile{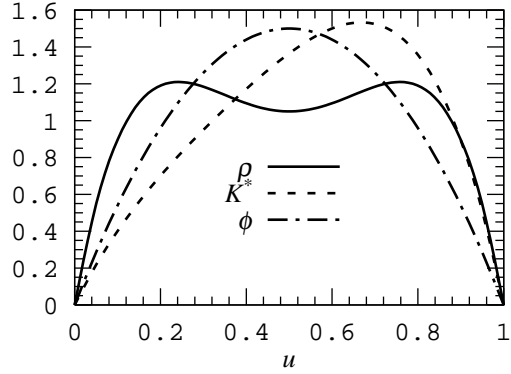}}
\end{figure}
\begin{table}
\renewcommand{\arraystretch}{1.4}
\addtolength{\arraycolsep}{3pt}
\caption[]{Masses and couplings of vector-meson distribution
  amplitudes, including SU(3)-breaking. In cases where two values 
are given, the upper one corresponds to the scale $\mu^2=1\,$GeV$^2$ 
and the lower one to $\mu^2 = 5\,$GeV$^2$,
respectively.
We use $m_s(1\,{\rm GeV}) = 150\,$MeV and put the $u$ and $d$ quark
mass to zero.
}\label{tab:para}
$$
\begin{array}{|c|cccc|}
\hline
V & \rho^\pm & K^*_{u,d} & \bar{K}^*_{u,d}& \phi\\ \hline
f_V [{\rm MeV}] & 198\pm 7  & 226 \pm 28& 226 \pm 28 & 254 \pm 3\\
f^T_V [{\rm MeV}] & 
\begin{array}{c} 160\pm 10 \\ 152\pm 9 \end{array}&
\begin{array}{c} 185\pm 10 \\ 175\pm 9 \end{array}&
\begin{array}{c} 185\pm 10 \\ 175\pm 9 \end{array}&
\begin{array}{c} 215\pm 15 \\ 204\pm 14 \end{array}
\\ \hline
a_1^\parallel & 0 & 
\begin{array}{c} 0.19 \pm 0.05 \\ 0.17\pm 0.04 \end{array}&
\begin{array}{c} -0.19 \pm 0.05 \\ -0.17\pm 0.04 \end{array}&
\phantom{-}0\\
a_2^\parallel & 
\begin{array}{c} 0.18 \pm 0.10 \\ 0.16\pm 0.09 \end{array}&
\begin{array}{c} 0.06 \pm 0.06 \\ 0.05\pm 0.05 \end{array}&
\begin{array}{c} \phantom{-}0.06 \pm 0.06 \\ 
                   \phantom{-}0.05\pm 0.05 \end{array}&
0\pm0.1\\
a_1^\perp & 0 & 
\begin{array}{c} 0.20 \pm 0.05 \\ 0.18\pm 0.05 \end{array}&
\begin{array}{c} -0.20 \pm 0.05 \\ -0.18\pm 0.05 \end{array}&
\phantom{-}0\\
a_2^\perp & 
\begin{array}{c} 0.20 \pm 0.10 \\ 0.17\pm 0.09 \end{array}&
\begin{array}{c} 0.04 \pm 0.04 \\ 0.03\pm 0.03 \end{array}&
\begin{array}{c} \phantom{-}0.04 \pm 0.04 \\ 
                  \phantom{-}0.03\pm 0.03 \end{array}&
 0\pm0.1\\ \hline
\delta_+ & 0 & 
\begin{array}{c} \phantom{-}0.24 \\ \phantom{-}0.22 \end{array}&
\begin{array}{c} 0.24 \\ 0.22 \end{array}&
\begin{array}{c} 0.46 \\ 0.41 \end{array}
\\
\delta_- & 0 & 
\begin{array}{c} -0.24 \\ -0.22 \end{array}&
\begin{array}{c} 0.24 \\ 0.22 \end{array}&
0 \\
\widetilde{\delta}_+ & 0 & 
\begin{array}{c} \phantom{-}0.16 \\ \phantom{-}0.13 \end{array}&
\begin{array}{c} 0.16 \\ 0.13 \end{array}&
\begin{array}{c} 0.33 \\ 0.27 \end{array}
\\
\widetilde{\delta}_- & 0 & 
\begin{array}{c} -0.16 \\ -0.13 \end{array}&
\begin{array}{c} 0.16 \\ 0.13 \end{array}&
0\\ \hline
\end{array}
$$
\renewcommand{\arraystretch}{1}
\addtolength{\arraycolsep}{-3pt}
\end{table}
\begin{table}
\caption[]{Couplings for twist-3 and 4 distribution amplitudes for
  which we do not include SU(3)-breaking. Renormalization scale as in
  the previous table.}\label{tab:para2}
\renewcommand{\arraystretch}{1.4}
\addtolength{\arraycolsep}{3pt}
$$
\begin{array}{|c|ccccccc|}\hline
& \zeta_3 & \omega_3^A & \omega_3^V & \omega_3^T & \zeta_4 & \zeta_4^T
& \tilde{\zeta_4^T}\\ \hline
V &
\begin{array}{c} 0.032\\ 0.023 \end{array}& 
\begin{array}{c} -2.1 \\  -1.8 \end{array}& 
\begin{array}{c} 3.8  \\   3.7 \end{array}& 
\begin{array}{c} 7.0  \\   7.5 \end{array}&
\begin{array}{c} 0.15 \\   0.13 \end{array}&  
\begin{array}{c} 0.10 \\   0.07 \end{array}&  
\begin{array}{c} -0.10\\  -0.07 \end{array}  
%-1.8 & 0.39 & 7.5 & 0.15 & 0.10 & -0.10
\\ \hline
\end{array}
$$
\renewcommand{\arraystretch}{1}
\addtolength{\arraycolsep}{-3pt}
\end{table}

\section{Summary of chiral-odd distributions}

There exist four different two-particle chiral-odd distributions \cite{BBKT}
defined as
\begin{eqnarray}
\lefteqn{\langle 0|\bar u(z) \sigma_{\mu \nu}  
d(-z)|\rho^-(P,\lambda)\rangle 
 = i f_{\rho}^{T} \left[ ( e^{(\lambda)}_{\perp \mu}p_\nu -
e^{(\lambda)}_{\perp \nu}p_\mu )
\int_{0}^{1} \!du\, e^{i \xi p \cdot z} \phi_{\perp}(u, \mu^{2}) \right. 
}\hspace*{3.7cm}\nonumber \\
& &{}+ (p_\mu z_\nu - p_\nu z_\mu )
\frac{e^{(\lambda)} \cdot z}{(p \cdot z)^{2}}
m_{\rho}^{2} 
\int_{0}^{1} \!du\, e^{i \xi p \cdot z} h_\parallel^{(t)} (u, \mu^{2}) 
\nonumber \\
& & \left.{}+ \frac{1}{2}
(e^{(\lambda)}_{\perp \mu} z_\nu -e^{(\lambda)}_{\perp \nu} z_\mu) 
\frac{m_{\rho}^{2}}{p \cdot z} 
\int_{0}^{1} \!du\, e^{i \xi p \cdot z} h_{3}(u, \mu^{2}) \right]\!\!,
\hspace*{0.5cm}\label{eq:tda}\\[-20pt]\nonumber\\ \nonumber
\end{eqnarray}
\begin{eqnarray}
\lefteqn{
\langle 0|\bar u(z)
d(-z)|\rho^-(P,\lambda)\rangle
 = }\hspace*{1.5cm}\nonumber\\
& & {} -i \left(f_{\rho}^{T} - f_{\rho}\frac{m_{u} + m_{d}}{m_{\rho}}
\right)(e^{(\lambda)}\cdot z) m_{\rho}^{2}
\int_{0}^{1} \!du\, e^{i \xi p \cdot z} h_\parallel^{(s)}(u, \mu^{2}).
\hspace*{0.4cm}\label{eq:sda}
\end{eqnarray}
The distribution amplitude $\phi_\perp$ is twist-2,
$h_\parallel^{(s)}$ and 
$h_\parallel^{(t)}$ are
twist-3 and $h_3$ is twist-4.   
All four functions $\phi=\{\phi_\perp,h_\parallel^{(s)},
h_\parallel^{(t)},h_3\}$ are normalized to
$$\int_0^1\!du\, \phi(u) =1.$$ 

Three-particle chiral-odd distributions are defined to twist-4 accuracy as
\begin{eqnarray}
\lefteqn{\langle 0|\bar u(z) \sigma_{\alpha\beta}
         gG_{\mu\nu}(vz) 
         d(-z)|\rho^-(P,\lambda)\rangle =}
 \nonumber \\
&=& f_{\rho}^T m_{\rho}^2 \frac{e^{(\lambda)}\cdot z }{2 (p \cdot z)}
    [ p_\alpha p_\mu g^\perp_{\beta\nu} 
     -p_\beta p_\mu g^\perp_{\alpha\nu} 
     -p_\alpha p_\nu g^\perp_{\beta\mu} 
     +p_\beta p_\nu g^\perp_{\alpha\mu} ] 
      {\cal T}(v,pz)
\nonumber\\
&+& f_{\rho}^T m_{\rho}^2
    [ p_\alpha e^{(\lambda)}_{\perp\mu}g^\perp_{\beta\nu}
     -p_\beta e^{(\lambda)}_{\perp\mu}g^\perp_{\alpha\nu}
     -p_\alpha e^{(\lambda)}_{\perp\nu}g^\perp_{\beta\mu}
     +p_\beta e^{(\lambda)}_{\perp\nu}g^\perp_{\alpha\mu} ]
      T_1^{(4)}(v,pz)
\nonumber\\
&+& f_{\rho}^T m_{\rho}^2
    [ p_\mu e^{(\lambda)}_{\perp\alpha}g^\perp_{\beta\nu}
     -p_\mu e^{(\lambda)}_{\perp\beta}g^\perp_{\alpha\nu}
     -p_\nu e^{(\lambda)}_{\perp\alpha}g^\perp_{\beta\mu}
     +p_\nu e^{(\lambda)}_{\perp\beta}g^\perp_{\alpha\mu} ]
      T_2^{(4)}(v,pz)
\nonumber\\
&+& \frac{f_{\rho}^T m_{\rho}^2}{pz}
    [ p_\alpha p_\mu e^{(\lambda)}_{\perp\beta}z_\nu
     -p_\beta p_\mu e^{(\lambda)}_{\perp\alpha}z_\nu
     -p_\alpha p_\nu e^{(\lambda)}_{\perp\beta}z_\mu
     +p_\beta p_\nu e^{(\lambda)}_{\perp\alpha}z_\mu ]
      T_3^{(4)}(v,pz)
\nonumber\\
&+& \frac{f_{\rho}^T m_{\rho}^2}{pz}
    [ p_\alpha p_\mu e^{(\lambda)}_{\perp\nu}z_\beta
     -p_\beta p_\mu e^{(\lambda)}_{\perp\nu}z_\alpha
     -p_\alpha p_\nu e^{(\lambda)}_{\perp\mu}z_\beta
     +p_\beta p_\nu e^{(\lambda)}_{\perp\mu}z_\alpha ]
      T_4^{(4)}(v,pz)\nonumber\\[-5pt]
\label{eq:T3}
\end{eqnarray}
and 
\begin{eqnarray}
\langle 0|\bar u(z)
         gG_{\mu\nu}(vz) 
         d(-z)|\rho^-(P,\lambda)\rangle 
&=& i f_{\rho}^T m_{\rho}^2
 [e^{(\lambda)}_{\perp\mu}p_\nu-e^{(\lambda)}_{\perp\nu}p_\mu] S(v,pz),
\nonumber\\
\langle 0|\bar u(z)
         ig\widetilde G_{\mu\nu}(vz)\gamma_5 
         d(-z)|\rho^-(P,\lambda)\rangle
&=& i f_{\rho}^T m_{\rho}^2
 [e^{(\lambda)}_{\perp\mu}p_\nu-e^{(\lambda)}_{\perp\nu}p_\mu] 
  \widetilde S(v,pz).
\end{eqnarray}
Of these seven amplitudes, ${\cal T}$ is twist-3 and the other six 
are twist-4. 

The light-cone expansion of the non-local tensor operator can be written
to twist-4 accuracy as
\begin{eqnarray}
\lefteqn{\hspace*{-1.5cm}\langle 0|\bar u(x) \sigma_{\mu \nu} 
d(-x)|\rho^-(P,\lambda)\rangle =} \nonumber \\
&=& i f_{\rho}^{T} \left[ (e^{(\lambda)}_{\mu}P_\nu -
e^{(\lambda)}_{\nu}P_\mu )
\int_{0}^{1} \!du\, e^{i \xi P x}
\Bigg[\phi_{\perp}(u) +\frac{m_\rho^2x^2}{4} {\Bbb A}_T(u)\Bigg] \right. 
\nonumber \\
& &{}+ (P_\mu x_\nu - P_\nu x_\mu )
\frac{e^{(\lambda)} \cdot x}{(P x)^{2}}
m_{\rho}^{2} 
\int_{0}^{1} \!du\, e^{i \xi P x} {\Bbb B}_T (u) 
\nonumber \\
& & \left.{}+ \frac{1}{2}
(e^{(\lambda)}_{ \mu} x_\nu -e^{(\lambda)}_{ \nu} x_\mu) 
\frac{m_{\rho}^{2}}{P  x} 
\int_{0}^{1} \!du\, e^{i \xi P x} {\Bbb C}_T(u) \right],
\label{eq:OPE2}
\end{eqnarray}
where ${\Bbb B}_T$ and ${\Bbb C}_T$ are expressed in terms of the distribution 
amplitudes defined above as 
\begin{eqnarray}
   {\Bbb B}_T(u) &=& h_\parallel^{(t)}(u) -\frac{1}{2}\phi_\perp(u)-
              \frac{1}{2} h_3(u),
\nonumber\\
   {\Bbb C}_T(u) &=& h_3(u)-\phi_\perp(u),
\end{eqnarray}
and ${\Bbb A}_T$ can be  related to integrals over three-particle distribution 
functions using the equations of motion.

%We introduce the notation, similar to Eq.~(\ref{eq:c12}):
%\begin{eqnarray}
% {\Bbb B}_T^{(i)}(u) &=& -\int_0^u\!dv\,{\Bbb B}_T(v),
%\nonumber\\
% {\Bbb C}_T^{(i)}(u) &=& -\int_0^u\!dv\,{\Bbb C}_T(v). 
%\end{eqnarray}
For the leading twist-2 distribution amplitude $\phi_\perp$ we use
\begin{equation}\label{eq:phiperp}
\phi_\perp(u) =  6 u\bar u \left[ 1 + 3 a_1^\perp\, \xi +
a_2^\perp\, \frac{3}{2} ( 5\xi^2  - 1 ) \right],
\end{equation}
with parameter values as specified in Table~\ref{tab:para}; the
definitions of $a_{1,2}^\perp$ are analaguous to Eqs.~(\ref{eq:a12}).
The expressions for higher-twist distributions given below correspond
to the simplest self-consistent approximation that satisfies all 
QCD equations of motion \cite{BBKT,BBS}:
\begin{itemize}
\item{} Three-particle distribution of twist-3:
\begin{equation}
{\cal T} (\underline{\alpha}) = 
540\, \zeta_3 \omega^T_3 (\alpha_d-\alpha_u)\alpha_d \alpha_u\alpha_g^2.
\end{equation}
\item{} Two-particle distributions of twist-3:
\begin{eqnarray}
h_\parallel^{(s)}(u) & = & 6u\bar u \left[ 1 + a_1^\perp \xi + 
\left( \frac{1}{4}a_2^\perp +
\frac{5}{8}\,\zeta_{3}\omega_3^T \right) (5\xi^2-1)\right]\nonumber\\
& & {}+ 3\, \delta_+\, (3 u \bar u + \bar u \ln \bar u + u \ln u) + 
3\,\delta_-\,  (\bar u
\ln \bar u - u \ln u),\label{eq:e}\\
h_\parallel^{(t)}(u) &= & 3\xi^2+ \frac{3}{2}\,a_1^\perp \,\xi (3 \xi^2-1)
+ \frac{3}{2} a_2^\perp\, \xi^2 \,(5\xi^2-3) 
\nonumber\\
& & {} +\frac{15}{16}\zeta_{3}\omega_3^T(3-30\xi^2+35\xi^4) 
+ \frac{3}{2}\,\delta_+
\, (1 + \xi \, \ln \bar u/u)\nonumber\\
&&{} + \frac{3}{2}\,\delta_- \, \xi\, ( 2
+ \ln u + \ln\bar u ).
\label{eq:hL}
\end{eqnarray}
\item{} Three-particle distributions of twist-4:
\begin{eqnarray}
 T^{(4)}_1(\underline{\alpha}) &=& T^{(4)}_3(\underline{\alpha}) = 0, 
 \nonumber\\
 T^{(4)}_2(\underline{\alpha}) &=& 
    30 \widetilde \zeta^T_{4}(\alpha_d-\alpha_u)\alpha_g^2,
 \nonumber\\
 T^{(4)}_4(\underline{\alpha}) &=& 
    - 30 \zeta^T_{4}(\alpha_d-\alpha_u)\alpha_g^2,
 \nonumber\\
 S(\underline{\alpha}) &=& 30 \zeta^T_{4}(1-\alpha_g)\alpha_g^2,
 \nonumber\\
 \widetilde S(\underline{\alpha}) &=& 
       30 \widetilde \zeta^T_{4}(1-\alpha_g)\alpha_g^2.
\end{eqnarray}
\item{} Two-particle distributions of twist-4:
\begin{eqnarray}
   h_3(u) &=& 6u(1-u)+5\left[
\zeta^T_4+\widetilde \zeta^T_4-\frac{3}{70} a_2^\perp\right](1-3\xi^2),
\nonumber\\
   {{\Bbb A}}_T(u) &=& 30 u^2(1-u)^2
   \Bigg[\frac{2}{5}+\frac{4}{35}a_2^\perp+\frac{4}{3}\zeta^T_4-\frac{8}{3}
   \widetilde \zeta^T_4\Bigg].
\end{eqnarray}
\end{itemize}
The constant $\omega_3^T$ is defined as
\begin{eqnarray}
\lefteqn{\langle 0|\bar u \sigma_{\mu\nu}z^\nu (gG^{\mu\beta} 
z_\beta (i\derright z) - (i\derleft z) g G^{\mu\beta}z_\beta) 
d |\rho^-(P,\lambda)\rangle  =}\hspace*{4cm}\nonumber\\
& = & (pz)^2 (e^{(\lambda)}z)
m_\rho^2 f_\rho^T \frac{3}{28}\, \zeta_3 \omega_3^T.
\end{eqnarray}
The constants $\zeta^T_4$ and $\widetilde \zeta^T_4$
are defined as
\begin{eqnarray}
\langle 0|\bar u gG_{\mu \nu}d|\rho^-(P,\lambda)\rangle &=&
  if_\rho^T m_\rho^3 \zeta^T_4(
e^{(\lambda)}_{\mu}P_\nu - e^{(\lambda)}_{\nu}P_\mu),
\nonumber\\
\langle 0|\bar u g\widetilde G_{\mu \nu}i\gamma_5 
   d|\rho^-(P,\lambda)\rangle &=&
  if_\rho^T m_\rho^3 \widetilde \zeta^T_4(
e^{(\lambda)}_{\mu}P_\nu - e^{(\lambda)}_{\nu}P_\mu).
\end{eqnarray}
Numerical values for all parameters are given in 
Table~\ref{tab:para}\footnote{In notations of Ref.~\cite{BBKT}
$\zeta_3^T \equiv (3/28)\zeta_3\omega_3^T$.}.
As in  the chiral-even case, we neglect SU(3)-breaking corrections
in twist-4 distributions. In Fig.~2, we plot the leading-twist
distribution amplitude $\phi_\perp$ for $\rho$, $K^*$ and $\phi$ mesons.
\begin{figure}
\caption[]{Twist-2 distribution amplitude $\phi_\perp(u,\mu=1\,{\rm
    GeV})$ for $\rho$, $K^*$ and $\phi$ mesons.}
\centerline{\epsffile{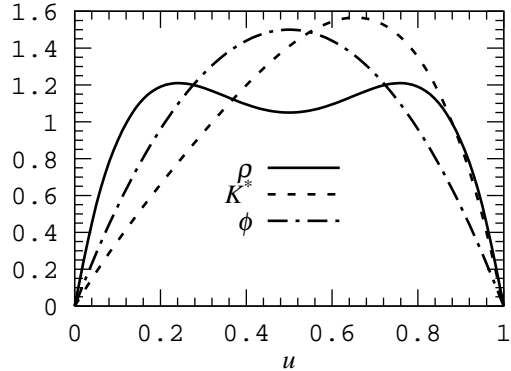}}\vspace*{-26pt}
\end{figure}
\section*{Acknowledgements}

We are grateful to Yu.\ Koike, G.\ Stoll and K.\ Tanaka for collaboration
at various stages of this project.
P.B.\ is supported by a Heisenberg-fellowship by DFG.

\end{document}